\documentclass[aps,pra,twocolumn,amsmath,amssymb,showpacs,superscriptaddress]{revtex4-1}
\usepackage{xcolor}
\usepackage{textcomp} 
\usepackage{mathrsfs,amsmath}
\usepackage{graphicx}
\usepackage{dcolumn}
\usepackage[mathlines]{lineno}
\usepackage{hyperref}
\usepackage{bm}
\usepackage{epstopdf}
\usepackage{soul}
\usepackage{epsfig}
\usepackage[normalem]{ulem}
\usepackage{bbold}
\usepackage{braket}
\usepackage{physics}
\usepackage{gensymb}
\usepackage{amsmath,amssymb}

\newcommand{\KD}[1]{\textcolor{orange}{#1}}

\begin{document}
\title{Orbital angular momentum of entangled photons as a probe for relativistic effects}% Force line breaks with \\

\author{Fazilah Nothlawala}
\affiliation{School of Physics, University of the Witwatersrand, Private Bag 3, Wits 2050, South Africa}

\author{Kiki Dekkers}
\email{kfhd2000@hw.ac.uk}
\affiliation{School of Engineering and Physical Sciences, Heriot-Watt University, Edinburgh, EH14 4AS, UK}

\author{Moslem Mahdavifar}
\affiliation{School of Physics, University of the Witwatersrand, Private Bag 3, Wits 2050, South Africa}

\author{Jonathan Leach}
\affiliation{School of Engineering and Physical Sciences, Heriot-Watt University, Edinburgh, EH14 4AS, UK}

\author{Andrew Forbes}
\affiliation{School of Physics, University of the Witwatersrand, Private Bag 3, Wits 2050, South Africa}

\author{Isaac Nape}
\email{isaac.nape@wits.ac.za}
\affiliation{School of Physics, University of the Witwatersrand, Private Bag 3, Wits 2050, South Africa}

%%%%%%%%abstract

%%%%%%%%abstract
\begin{abstract}
\noindent Orbital angular momentum (OAM) as both classical and quantum states of light has proven essential in numerous applications, from high-capacity information transfer to enhanced precision and accuracy in metrology.  Here, we extend OAM metrology to relativistic scenarios to determine the Lorentz factor of a moving reference frame, exploiting the fact that OAM is not Lorentz invariant. We show that the joint OAM spectrum from entangled states is modified by length contraction when measured by two observers moving relative to the entanglement source. This relative motion rescales the spatial dimensions, thus breaking the orthogonality of the OAM measurement process and resulting in a broadening of the joint OAM spectrum that can precisely determine the Lorentz factor. We experimentally simulate velocities up to $0.99c$, confirm the predicted broadening, and use the measurement outcomes to extract the Lorentz factor. Our work provides a pathway for novel measurement techniques suitable for relativistic conditions that leverage OAM structured light as a resource.
\end{abstract}

\maketitle
\section{Introduction}
\noindent Light with $\ell \times 2 \pi$ helical twists to its phase is endowed with orbital angular momentum (OAM) \cite{allen1992orbital}, enabling the direct rotation of objects via structured wavefronts \cite{he1995direct} as bright classical beams, and  $\ell \hbar$ of OAM per photon as quantum states \cite{mair2001entanglement}. These traits have fueled many applications and fundamental insights that have been extensively reviewed to date \cite{barnett2017optical,erhard2018twisted,forbes2024orbital,willner2015optical,padgett2017orbital,franke202230,yang2021optical,forbes2021structured}. A recent surge is in OAM‐based metrology and sensing \cite{cheng2025metrology}, where OAM has emerged as an essential resource for ultra-sensitive measurements. It has enabled super‐resolution imaging \cite{li2013beating, shi2023super}, measurement of uniform or rough rotating surfaces \cite{lavery2013detection, wan2025compact, ren2022non}, and for novel magnetic‐field detection via the Faraday effect \cite{pang2019orbital}. All these applications exploit the OAM-carrying beam's response to physical motion. For example, the lateral displacements of an object can alter the OAM spectrum of a probing beam \cite{cvijetic2015detecting} and the change in temperature (velocity field) of a medium can affect the interference of OAM modes \cite{zhang2021tiny}. Similarly, the rotation of an object can induce a detectable rotational Doppler effect, enabling one to retrieve the rotational velocity of physical objects \cite{xie2017using, ren2022non}, and can be extended to translation and rotation by employing OAM in a vectorial combination \cite{fang2021vectorial}.  %In some cases, the physical effects can allow for simultaneous detection of rotational and translational motion \cite{lu2025angular}. 

The above mentioned examples illustrate applications in the classical regime. With the increasing range of tools for generating and detecting single and entangled photons carrying OAM \cite{nape2023quantum, gao2024full, d2019tunable, agnew2014discriminating, wu2022room}, metrology using these states has advanced into the single‐photon regime. Example applications include photonic gears for angular measurements using single and two photon OAM states \cite{d2013photonic, yesharim2024quantum}  as well as digital spiral imaging at the single‐photon level, which exploits the OAM spectrum of entangled fields to measure a distant object's phase nonlocally \cite{molina-terriza2007probing, chen2014quantum}. The latter has also underpinned ghost‐diffraction experiments, where a photon’s OAM scattering is inferred from its entangled partner \cite{chen2010high}, and has stimulated discussions on uncertainty relations and nonlocality in the angular‐momentum degree of freedom \cite{jha2008fourier, gounden2023popper}.

An intriguing frontier considers optical angular momentum in special and general relativity. Because a photon’s OAM is not Lorentz invariant \cite{bliokh2012spatiotemporal}, probing it at extreme velocities reveals novel uncertainty relations and apparent paradoxes \cite{leader2014angular, barnett1994orbital, tamburini2020relativistic}. On the other hand, because photons do not have a rest frame, it has been highlighted that spin and orbital angular momentum are coupled \cite{van1994commutation, van1994spin, bliokh2010angular} in light fields, yet the separation is possible if the angular momentum is projected onto its longitudinal components. Remarkably, under Lorentz boosts, the longitudinal OAM components can also be decomposed into extrinsic and intrinsic parts \cite{bliokh2012spatiotemporal}, giving rise to relativistic Hall effects in optical vortices \cite{bliokh2012relativistic}. Nonetheless, the longitudinal OAM states are physical and have been measured extensively as is seen in earlier examples.  Recent proposals even suggest using OAM modes as remote sensors of black‐hole rotation,  opening up the use of OAM to a wider range of physical scenarios \cite{tamburini2022constraining, tamburini2011twisting}, with gravitational effects realizable in a laboratory setting \cite{rodriguez2023einstein}. Central to these studies is the characterisation of the motion via the OAM spectrum, specifically by exploiting its broadening under the effect of gravitational lensing. To this end, this opens up the question of whether the OAM spectrum of single photons can be used as a probe to measure physical quantities in relativistic reference frames.

\begin{figure*}[t!]
    \centering
    \includegraphics[width=1\textwidth]{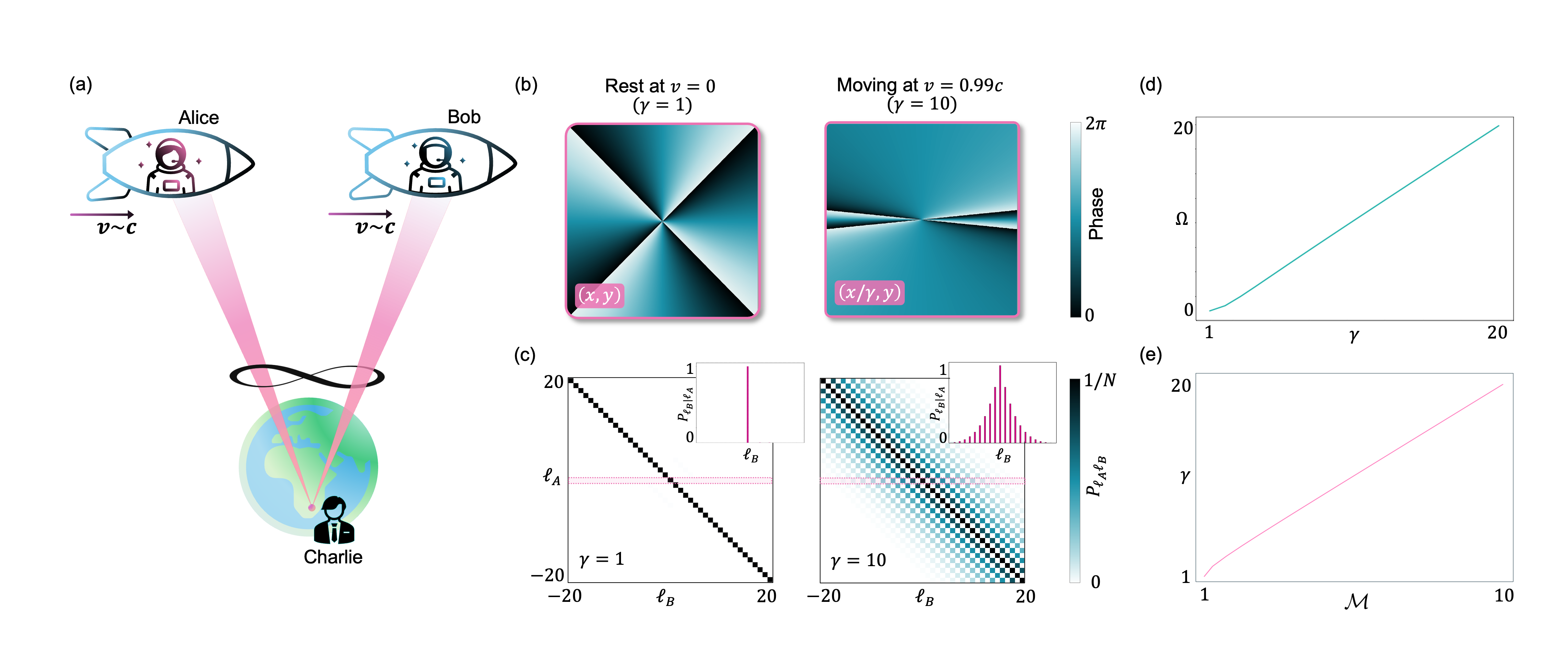}
    \caption{(a) Two photons entangled in their OAM degree of freedom are sent to two independent detectors, Alice (A) and Bob (B), that are moving at a speed $v \sim c$. In Charlie's reference frame, this is seen as a length contraction in the x-direction of the detectors, mapping the coordinates $(x, y) \rightarrow (x/ \gamma, y)$.  The detectors project onto the OAM eigenstates $\ket{\ell_A}$ and $\ket{\ell_B}$, where Charlie observes their OAM phase patterns as shown in (b) for Alice and Bob at rest ($v = 0, \gamma=1$) or moving at relativistic speeds ($v = 0.99c, \gamma=10$). The resulting OAM correlations are shown in (c) for $\gamma = 1$ and 10 with insets indicating the conditional probability, $P_{\ell_B| \ell_A = 0 }$, in the range $\ell_B =  \{-20, 20\}$. For $\gamma =1$ the plot shows that the joint detection probability has OAM anti-correlations, i.e. non-zero detection probability for $\ell_A=-\ell_B$. (d) However, as $\gamma$ increases, the orthogonality is compromised and the number of contributing modes, $\Omega$ increases with the increment of the Lorentz factor. (e) Consequently, the Lorentz factor can be extracted from the observed OAM spectrum using the measurement probabilities $\mathcal{M}$ computed from Eq.~(\ref{eqn:gammaextract}) which is monotonic and near linear with respect to $\gamma$.}
    \label{fig:concept}
\end{figure*}

Here, we demonstrate that the broadening in OAM due to Lorentz boosts can determine physical parameters from special relativity. We show that the technique can be used to estimate the Lorentz factor, $\gamma$, from a pair of synchronised relativistic inertial reference frames. In such frames, length contraction rescales spatial coordinates of detectors and distorts the orthogonality of measured entangled photons, producing a measurable broadening in detected OAM  correlations. For two photons entangled in OAM, this effect modifies their joint correlations, enabling direct inference of $\gamma$ from the altered distribution. To validate our approach, we generate entangled photon pairs via SPDC and emulate relativistic length contraction with spatial light modulators. By comparing the measured OAM spectra against theoretical predictions, we demonstrate quantitative agreement and establish structured light as a novel metrology tool for probing relativistic effects.

\section{Theory}
\subsection{Concept}
Consider Alice (A) and Bob (B) moving at relativistic speeds with respect to Charlie (C), who is in a rest frame, as depicted in Fig.~\ref{fig:concept} (a). Charlie sends one photon from an entangled photon pair to Alice and the other to Bob. The aim is to use the entangled photons to investigate the relativistic motion of Alice and Bob via their correlated measurements. To begin, we assume that the photons are entangled in the orbital angular momentum (OAM) degree of freedom and are described by the two-photon state, $\ket{\Psi} = \frac{1}{ \sqrt{N}} \sum_{\ell} \ket{\ell}_A\ket{-\ell}_B$, where $N$ is the number of modes contributing to the quantum state. The OAM eigenstates $\ket{\ell} \propto \int \exp(i \ell \phi ) \ket{\phi} d \phi$ are characterised by an azimuth  dependent  ($\phi = \atan(y/x)$) phase profile, $ \exp(i \ell \phi )$, with $\ell$ corresponding to an OAM of $\ell\hbar$ per photon. 

%Suppose that the two photons are measured in the inertial reference frames of Alice and Bob who are moving at the same velocity ($v$) as shown in Fig. \ref{fig:concept} (a). Accordingly, we can characterise their motion using the Lorentz factor $\gamma = 1/{\sqrt{1- \frac{v^2}{c^2}}}$, where the velocity vector aligns with one of the coordinate axes of the measurement system (i.e. the x-direction).  Consequently, Charlie observes Alice and Bob's measurement devices to be contracted and describes them using transformed spatial coordinates, $(x, y) \rightarrow (\frac{x}{\gamma}, y)$, scaling the x-coordinate by a factor $\gamma$ due to length contraction (See Appendix A).
%To capture the relativistic effects on the photon detectors, we describe them here from the perspective of a rest-frame observer (Charlie), who sees the detectors as Lorentz-contracted. However, we emphasise that the detection probabilities equate the same when evaluated in the boosted reference frames (Alice and Bob's), who observe the incoming photons as sheared in the $x$ direction (see supplementary material). Further, because Alice and Bob experience the same time dilation, we assume that they remain synchronised.

To describe Alice and Bob's measurement states while moving at relativistic speeds, we use the Lorentz factor $\gamma = 1/{\sqrt{1- \frac{v^2}{c^2}}}$, where the velocity vector aligns with one of the coordinate axes of the measurement system (i.e., the $x$-direction). As a result,  Alice and Bob's measurements are altered so that when viewed from a rest reference frame, their transverse coordinates are mapped via $(x, y) \rightarrow (\frac{x}{\gamma}, y)$, scaling the $x$-coordinate by a factor $\gamma$ due to length contraction (see Supplementary Material).  Crucially, because Alice and Bob experience the same time dilation, we assume that they remain synchronised.

It is important to note that it does not matter whether this scenario is viewed from Alice and Bob's or Charlie's reference frames. Either Charlie is at rest and observes the measurement apparatus to be length-contracted, or Alice and Bob are at rest and they observe the two-photon state to be length-contracted. In this work, we mainly explore this scenario from the former perspective, but we highlight that the results would be equivalent when considered from the latter perspective.

\subsection{$\gamma$-dependent OAM Measurement Probabilities}
The two-photon state can be rewritten in the azimuthal coordinate basis as $\ket{\Psi} \propto \int \ket{\phi} \ket{\phi} d \phi $, where the inner product relation $\langle \phi|\phi' \rangle = \delta(\phi' - \phi)$ holds (See Supplementary for the mapping). This description of the state can be obtained by applying the resolution of the identity to the OAM entangled states using the continuous azimuthal ($\ket{\phi}$) basis, i.e., $\int_{0}^{2\pi} d\phi \; \ket{\phi}\bra{\phi} = \mathbb{I}$, in its continuous form, or alternatively by expressing OAM states in the continuous azimuth basis as we have shown in the Supplementary information. By defining the two-photon states in this way, we see that the entangled state acts as a projector that maps Alice's and Bob's states onto wavefunctions described in terms of the $\phi$ coordinate as seen from the rest frame. This means that the projection states used by Alice and Bob take the form $\langle  \psi_{A, B}  | \phi \rangle = \psi^{*}_{A(B)}(\phi)$, defining the conjugate of the wavefunctions of Alice or Bob measurements in terms an observer at rest (Charlie).  For this reason, we will describe the projection states using Charlie's rest frame coordinates from here on.\\

The OAM projections that Alice and Bob perform are described using length-contracted azimuthal coordinates that are transformed as $\phi~\rightarrow~\phi'~(\phi)~=~\arctan (\gamma \tan(\phi))$, resulting in measurements corresponding to $ \ket{\ell_{A, B}}~\propto~\int \exp(i \ell_{A,B} \phi'(\phi)) \ket{\phi} d \phi.$
In Fig.~\ref{fig:concept}~(b) we illustrate the effect of length contraction on the OAM measurement patterns for Alice and Bob at rest ($v = 0$) and Alice and Bob moving at a velocity close to the speed of light ($v = 0.99c$). Using these phase profiles, the joint detection probabilities can be obtained from solving the overlap integral given by 
\begin{align}
   P_{\ell_A \ell_{B}} =& | \langle \ell_B | \langle \ell_A | \psi \rangle|^2 \nonumber, \\ 
   =&  \left| \frac{1}{\sqrt{N}}\int_{0}^{2 \pi} \langle \ell_A | \phi \rangle \  \langle \ell_B| \phi \rangle \ d\phi \nonumber \right| ^2, \\
   =& \left| \frac{1}{2\pi\sqrt{N}}\int_{0}^{2 \pi} \exp\left(-i \left(\ell_A +\ell_B\right)\phi'\left(\phi\right)\right) \ d\phi \right|^2, \nonumber \\ 
   =& \left |  \frac{1}{2\pi \sqrt{N}} \int_{0}^{2 \pi}   \frac{\gamma \text{ exp} (-i (\ell_A + \ell_B) \phi'(\phi))}{(\gamma^2-1)\cos^2(\phi') +1}   d \phi'  \right |^2,
   \label{eq:overlapint}
\end{align}
\noindent where the azimuthal functions arise from the distorted azimuthal coordinates $\phi' = \text{arctan}(\gamma\text{tan}(\phi))$, with the appropriate mapping for azimuth differential element applied following
\begin{equation}
d\phi = \frac{\gamma}{(\gamma^2-1)\cos^2 \phi' + 1'} d\phi'.
\end{equation}

Importantly, an identical integral is obtained when considering the computation from Alice and Bob's reference frames. From their perspective, Alice and Bob observe the photons as being contracted, and it appears as though the photon states now undergo distortion. In that case, one solves the integral by applying a Lorentz boost to the Cartesian coordinate system of the joint two-photon wavefunction. This transforms the coordinates as, $(x', y') \rightarrow (\gamma x', y')$, assuming we have set $t'=0$, resulting in  
\begin{align}
    P_{\ell_A\ell_B} &\propto  \Big| 
    \int_{0}^{2\pi} \int_{0}^{\infty} 
    \underbrace{\exp\!\left( -r'^2 (\gamma^2 \cos^2\phi' + \sin^2\phi')\right)}_{\text{entangled state}} \notag\\ 
    & \quad \times 
    \underbrace{\exp(-i\ell_A \phi')}_{\text{Alice}}
    \underbrace{\exp(-i\ell_B \phi')}_{\text{Bob}} \,
    r' dr' d\phi' \Big|^2, \\
    &= \Big| \int_{0}^{2\pi} \int_{0}^{\infty} 
    \exp \!\left( -r'^2 \big((\gamma^2-1)\cos^2\phi' +1\big) \right) \notag \\ 
    & \quad \times 
    \exp \!\left(-i(\ell_A+\ell_B) \phi' \right)\,
    r' dr' d\phi' \Big|^2 .
\end{align}
This is an adaptation of the two-photon integral for SPDC, assuming that a Gaussian function defines the joint state in the near field, \KD{$\psi$} $\propto\exp(-r^2)$, written in normalised radial coordinates ``$r$" in Charlie's reference frame. From Alice and Bob's perspective, the two-photon state appears to be distorted due to their motion, such that the radial coordinate of the photons is mapped as $ r = \sqrt{r'^2 (\gamma^2 \cos^2(\phi') + \sin^2(\phi'))} = r'\sqrt{ ((\gamma^2-1)\cos^2(\phi') +1)}$. Next, we use the fact that 
\begin{equation}
    \frac{dr}{dr'} = \sqrt{(\gamma^2-1)\cos^2(\phi') +1}, 
\end{equation}
transforming the integral to
\begin{equation}
    P_{\ell_A \ell_B} \propto \left| \int_{0}^{2\pi} \int_{0}^{\infty} \exp \left(- r^2 \right) \frac{ \exp \left(-i(\ell_A+\ell_B) \phi' \right)}{(\gamma^2-1)\cos^2(\phi') +1} r dr d\phi' \right|^2 
\end{equation}
The radial part of the integral evaluates to a constant, while the azimuthal part will result in the same integral as Eq. (\ref{eq:overlapint}). This confirms that the probability outcomes are independent of whether the computation is done in the moving frames (Alice and Bob) or the stationary reference frame (Charlie).

Upon solving the complex integral \KD{\eqref{eq:overlapint}} (See Supplementary), one obtains a form solution
 \begin{equation}
   P_{\ell_{A}\ell_B} =  \frac{1}{N}  \left( \frac{\gamma -1 }{\gamma+1} \right)^{|\ell_{A}+\ell_{B}|} \cos^2 \left( (\ell_A+\ell_B)\frac{\pi}{2}\right), 
   \label{eq: solved joint probability}
\end{equation}
given a measurement of $\ell_A$ and  $\ell_B$ by Alice and Bob.  

Note that the orthogonality between Alice and Bob's measurement projection states is broken, such that their measurement scheme no longer constitutes a positive operator-valued measure (POVM). Thus, the sum of the joint probabilities over all possible  $\ket{\ell_A}$ and $\ket{\ell_B}$ measurements can now exceed one.
The corresponding theoretical joint probability spectra are shown in Fig.~\ref{fig:concept} (c).
The $\cos^2 \left( \left(\ell_A+\ell_B \right)\frac{\pi}{2}\right)$ modulation restricts the correlations to only exist when the sum of $\ell_A$ and $\ell_B$ is an even integer ($\ell_A + \ell_B \equiv \text{even}$). This is confirmed in the inset of each probability spectrum, where the conditional probabilities are shown for Alice projecting on $\ell_A = 0$ while Bob's projection measurements range over  $\ell_B =~\{-20, 20 \}$. 

Further, for any value of $\ell_A$, the broadening in OAM ($\ell_B$) is identical. Therefore, for any given $\gamma$,  it can be seen that the distribution in the OAM that Bob observes is independent of the OAM that Alice measures, i.e.,~it only depends on the sum $\ell_{A,B} = \ell_A+\ell_B$.  

%We can quantify the effective number of contributing modes, defined as $\Omega = \frac{ \left( \sum_{\ell_B} P_{\ell_B|\ell_A}\right)^2 }{  \sum_{\ell_B} P^2_{\ell_B|\ell_A}}$  for any value of $\ell_A$ ~\cite{law2004analysis},  which increases monotonically with $ \gamma \geq 1 $, as shown in Fig.~\ref{fig:concept}(d). 
%This expression was computed analytically, 
%$\Omega = \frac{(1 + \gamma^2)^3}{\gamma(1 + 6\gamma^2 + \gamma^4)} $,  providing a direct estimate of the distribution’s effective width, therefore establishing a one-to-one correspondence between $ \gamma $ and the modal spread (see Supplementary for details on the derivation).
\subsection{Broadening of Conditional Probability}
Since the detection probabilities are a function of $\ell_A~+~\ell_B$, we study the spread given a Lorentz factor by fixing $\ell_A$ and studying the detection probabilities with respect to $\ell_B$. Therefore, the conditional probabilities become 
\begin{equation}
   P_{\ell_B|\ell_A} =  \left( \frac{\gamma - 1 }{\gamma+1} \right)^{|\ell_A+\ell_B|} \cos^2(\left(\ell_A+\ell_B\right) \frac{\pi}{2}). 
\end{equation}
Therefore, the effective number of contributing modes is quantified using the formula
\begin{align}
    \Omega_B &= \frac{ \left( \sum_{\ell_B} P_{\ell_B|\ell_A}\right)^2 }{  \sum_{\ell_B} P^2_{\ell_B|\ell_A}}, \nonumber \\
&= \frac{(1+ \gamma^2)^3}{\gamma(1+6\gamma^2 + \gamma^4)},
\end{align}
and similarly for $\Omega_A.$ This is a monotonically increasing function with $ \gamma \geq 1 $, as shown in Fig.~\ref{fig:concept}(d). 
This expression was computed analytically,  providing a direct estimate of the distribution’s effective width, therefore establishing a one-to-one correspondence between $ \gamma $ and the modal spread. Such a relation is commonly used to estimate the number of effective modes in a given distribution \cite{law2004analysis}. For example, when $\gamma = 1$, the expression yields $\Omega_{B} =1$, meaning that Bob detects a spectrum that effectively has one contributing mode when Alice projects onto any given mode. However, for $\gamma=10$, Bob will see a spectrum that, on average, has $\Omega_{A,B}  \approx 10$ modes. Alternatively, the spread can be characterised via the standard definition of uncertainty, $\sqrt{ \langle  \ell^2_B \rangle - \langle \ell_B \rangle^{2}}$,  yielding the expression  $ \frac{\gamma - \gamma^{-1}}{\sqrt{2}} $, when the measurement detection events are normalised, capturing the width observed by Bob when Alice projects onto a single OAM mode.

\subsection{Extracting the Lorentz factor $\gamma$}
Next, we show how we can extract the Lorentz factor using the conditional probability, $P_{\ell_B|\ell_A}=N\times P_{\ell_A\ell_B}$.

%$ \Delta \ell_{{A,\, B}} \propto \sqrt{ \langle \ell_{A,\, B}^2 \rangle  - \langle \ell_{{A,\, B}} \rangle^2}$

First, for any projection on $\ell_A$, the summed conditional probabilities, $\sum_{\ell_B}P_{\ell_B|\ell_A}$ over $\ell_B$ is equal to
\begin{align}
\mathcal{M}  &=  \sum^{\infty}_{\ell_B=-\infty} \left( \frac{\gamma -1 }{\gamma+1} \right)^{|\ell_A + \ell_B|} \cos^2(|\ell_A + \ell_B|\frac{\pi}{2}) \\
 &=  \frac{\gamma+\gamma^{-1}}{2},
\label{eq:sumprob}
\end{align}
and is independent of Alice's projected state. This means that the sum of the conditional probabilities over Bob's measurements is scaled by $\mathcal {M} =\frac{\gamma+\gamma^{-1}}{2}$, and $\mathcal{M}$ can now take on values above 1, arising from the non-orthogonal nature of the distorted measurements and overlapping measurement contributions.

%\sout{Further, given that Alice projects her photon onto the state $\ell_{A}$, the total OAM that Bob measures, given that Alice projected her photon onto the state $\ell_{A}$, may appear to be given by $\langle \ell_B   \rangle = \sum_{\ell_B} P_{\ell_A|\ell_B} \ell_B  \propto -\frac{\gamma+\gamma^{-1}}{2} \ell_{A}$, where the OAM is scaled by the factor $\mathcal{M} = \frac{\gamma+\gamma^{-1}}{2}$.  Interestingly, this is the same factor that has been identified in Ref. \cite{bliokh2012spatiotemporal} , which found that exactly this amount modifies the total intrinsic OAM due to relativistic boosts on vortex patterns. }

Second, the total OAM that Bob measures given Alice’s measurement, would be given by $\langle\ell_B\rangle~=~\sum_{\ell_B}~P_{\ell_A|\ell_B} \ell_B = -\frac{\gamma+\gamma^{-1}}{2} \ell_{A}$.  Here we see Bob's OAM is scaled by the factor $\mathcal{M} = \frac{\gamma+\gamma^{-1}}{2}$, which is the same factor that modifies the total intrinsic OAM due to relativistic boosts on vortex patterns, as identified in Ref. \cite{bliokh2012spatiotemporal}.  This scaling of the OAM appears when using the un-normalised probabilities, $P_{\ell_A|\ell_B}$, and in our work, this merely emerges as a correction factor that disappears when the detection outcomes are correctly normalised. Once the correction is applied, the observable on Bob's OAM becomes $\langle \ell_B \rangle =- \ell_A$, which is corroborated by the symmetric OAM spectrum that Bob observes and is centered at~$- \ell_A$.    
% Accounting for the extrinsic OAM, the total OAM transforms as $L^{(\text{int.})} +  L^{(\text{ext.})} =\gamma L$ \cite{bliokh2012spatiotemporal}. 

Finally, we see that knowledge of $\mathcal{M}$ establishes the relative motion between the detectors by solving for $\gamma~\geq~1$,
\begin{equation}
\gamma =\mathcal M  + \sqrt{\mathcal {M}^{2} -1},
\label{eqn:gammaextract}
\end{equation}
from which Alice and Bob can estimate the Lorentz factor, as illustrated in Fig. \ref{fig:concept} (e). Moreover, for large $\gamma $, $\mathcal{M} \approx \gamma/2$, making it possible to estimate the Lorentz factor directly from $\mathcal{M}$.\\

% , provided the detectors are well-calibrated and all the noise sources are characterised.

% Because Lorentz transformations can be treated as hyperbolic rotations in space-time coordinates, where the mapping between $(t, x)$  and $ (t', x')$ is achieved through a rotation in Euclidean space by an angle $\eta$ (the rapidity), following $\begin{pmatrix}
% \cosh(\eta) & -\sinh(\eta)\\
% \sinh(\eta) & \cosh(\eta)
% \end{pmatrix} \begin{pmatrix}
% ct \\ x 
% \end{pmatrix} =\begin{pmatrix}
% ct' \\ x' 
% \end{pmatrix}$, where $\cosh(\eta) = \gamma$. Therefore, we can equate
% \begin{equation}
%    C_{\ell_A \ell_B} =  \frac{1}{N}  \left( \tanh \left( \frac{\eta}{2}\right)^{ |\ell_{A}+\ell_{B}|} \cos\left((\ell_A+\ell_B)\frac{\pi}{2}\right) \right)^2, 
%    \label{eqn:eta_fit}
% \end{equation}
% and so Alice and Bob can use the spectrum to fit $\eta$, which can take into account noise or other defects as part of the fitting routine. %
\begin{figure}[t]
    \centering
\includegraphics[width=1\linewidth]{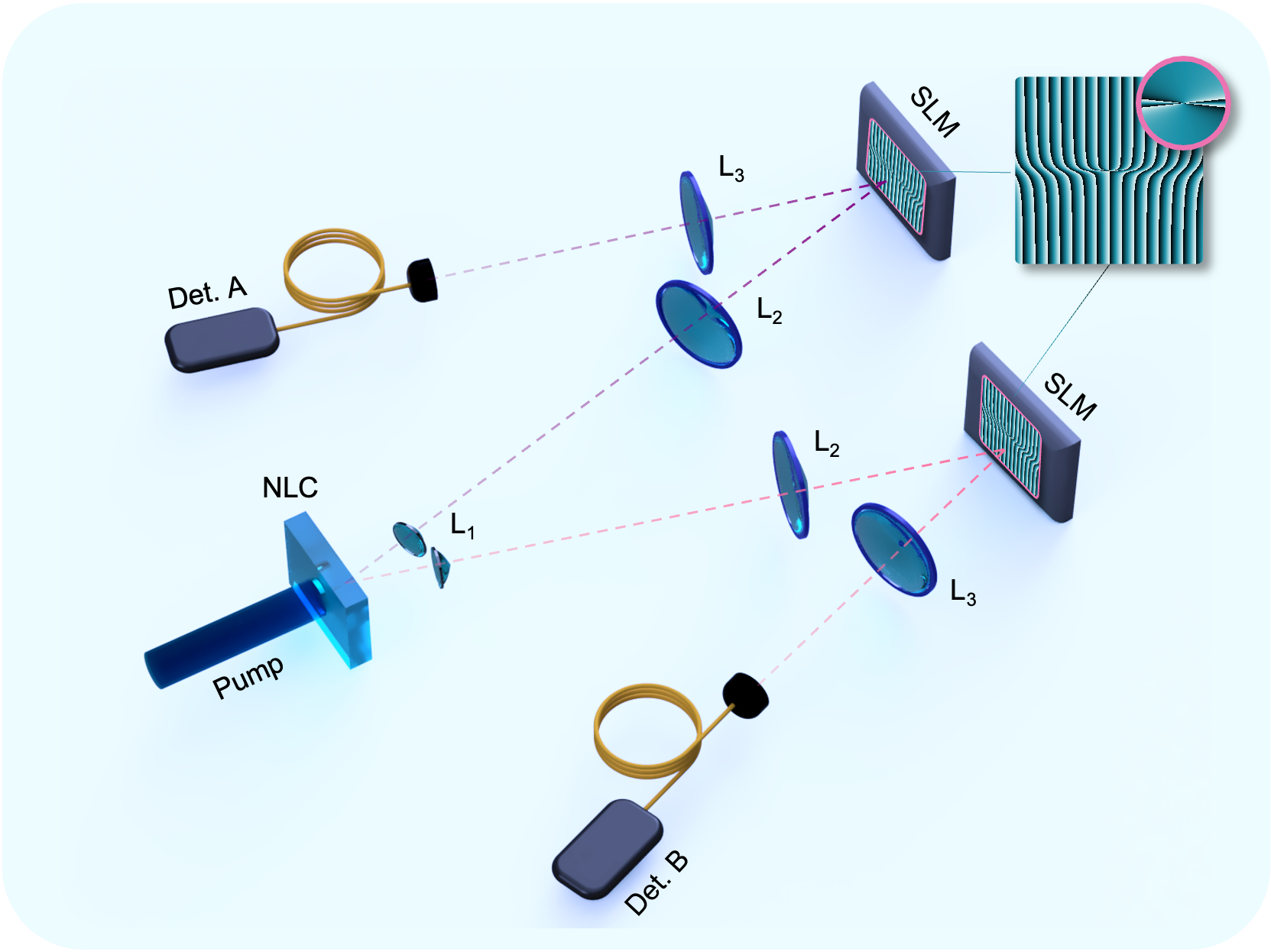}
    \caption{Schematic diagram of the non-degenerate quantum experiment used to create OAM entangled photons. A non-linear crystal (NLC) is used to produce dual-wavelength SPDC at $\lambda_1$ = 810 nm and $\lambda_2$ = 1550 nm. The plane of the crystal is imaged onto two spatial light modulators (SLM). OAM projection holograms are displayed on each SLM with the $x$-coordinate contracted as $x\rightarrow x/\gamma$ to perform joint projections. The photons are thereafter coupled into SMFs connected to single-photon detectors (Det. A(B)).}
    \label{fig:ExpSetup}
\end{figure}

\section{Experiment}
%\section{Experiment and results}
To emulate these predicted relativistic effects, we used photons generated using nondegenerate spontaneous parametric down conversion (SPDC) as shown in Fig. \ref{fig:ExpSetup}. To achieve this, a pump beam, of wavelength $\lambda_p = 532 $ nm, was used to generate collinear, vertically polarised photon pairs at wavelengths $\lambda_1 = 810$ nm and $\lambda_2 = 1550$ nm using a type-I temperature‐controlled ppKTP crystal. The two photons were imaged onto independent spatial light modulators (SLMs),i.e., PLUTO-NIR and PLUTO-TELCO SLMs for OAM projections, and thereafter coupled into single‐mode fibers and subsequently detected using APDs tailored for the individual wavelengths.  The joint probability spectra were obtained by performing OAM projections using the SLMs and single-mode fibers that together served as our effective, contracted detectors. Importantly, the distorted detection modes were encoded as phase holograms (see inset) that have a contracted  $x$-coordinate (i.e. $(\frac{x}{\gamma}, y)$). Subsequently, the resulting modulated photons were collected with single-mode fibers and detected in coincidence.

\section{Results and Discussion}
\begin{figure*} [t]
    \centering
     \includegraphics[width=1\textwidth]{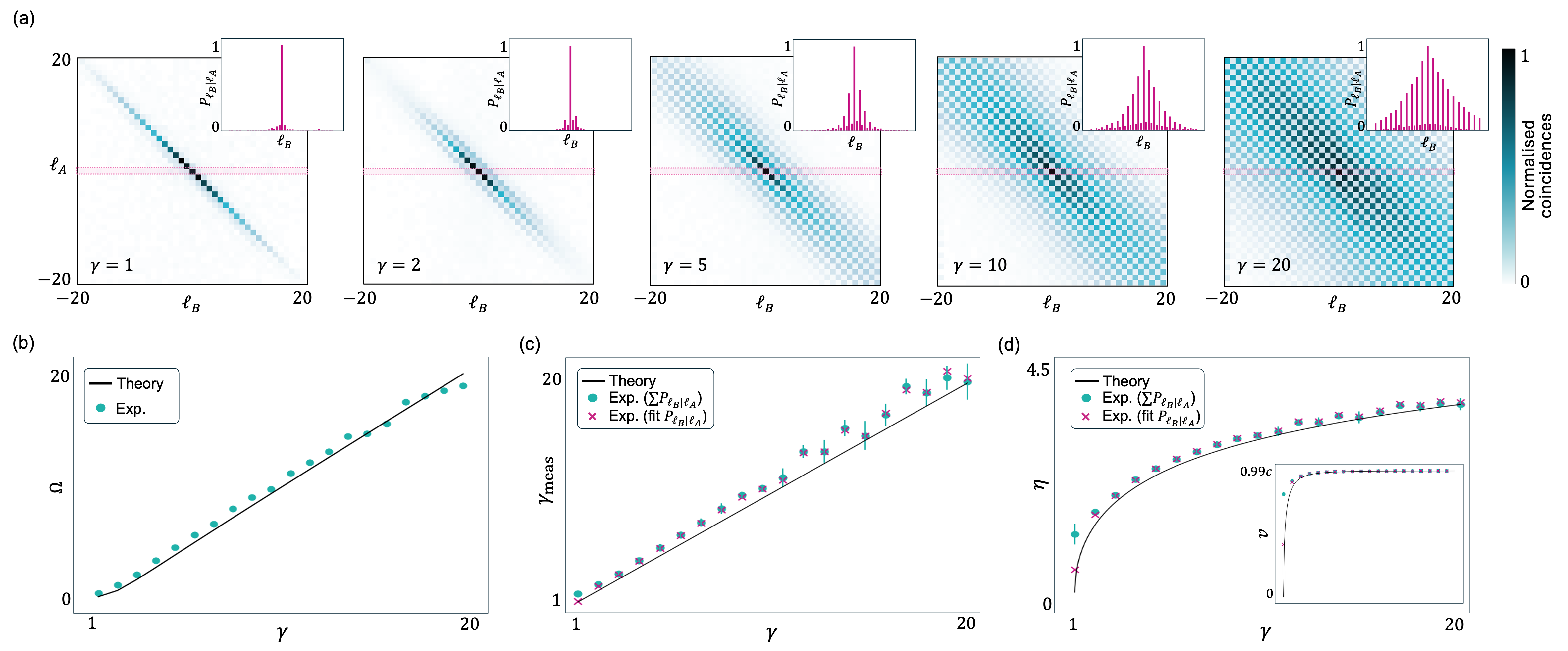}
    \caption{(a) Measured coincidences for encoded $\gamma=$ 1, 2, 5, 10 and 20. Insets show the conditional probability, $P_{\ell_B| \ell_A = 0 }$, for a fixed $\ell_A =0$ in the range $\ell_B = \{-20, 20\}$.
    (b) Measured $\Omega$ (number of contributing modes) vs the encoded Lorentz factors $\gamma$. (c) The encoded Lorentz factor $\gamma$ compared with the measured factor, $\gamma_\text{meas}$. The fitted $\gamma$ values using Eq. \ref{eq:sumprob} are represented by pink crosses. (d) A plot for the related rapidity $\eta$, computed using the experimentally measured Lorentz factor in comparison with theoretical $\eta$. The inset represents the velocities determined from $\gamma_\text{meas}$, showing simulated velocities up to $0.99c$. }
    \label{fig:EXP}
\end{figure*}

The experimentally measured joint probability spectra for encoded $\gamma = 1, 2, 5, 10$, and 20 are presented in Fig. \ref{fig:EXP}(a). These results align well with the theoretical joint probabilities shown in 
Fig. \ref{fig:concept}(c), across the range of $\gamma$ values. 
For each case, a cross-section of the spectrum at $\ell_A = 0$ is shown as an inset, where the conditional probability amplitudes have been normalised by the peak. 
In Fig. \ref{fig:EXP} (b), we observe that the width of the OAM spectrum broadens with increasing $\gamma$ as expected. 
We extract the measured $\gamma_\text{meas}$ through Eq. \eqref{eqn:gammaextract} by summing the detected probabilities $P_{\ell_B| \ell_A = 0 }$ for even $\ell_B$ (corresponding to $\ell_A + \ell_B \equiv \text{even}$) to obtain $\mathcal{M}$. Additionally, to reduce the impact of experimental noise, we fitted the theoretical conditional probability, related to Eq. \eqref{eq: solved joint probability} via $P_{\ell_B|\ell_A}=N\times P_{\ell_A\ell_B}$, to these detected probabilities to find the best fit parameter for $\gamma_\text{meas}$. 
% \sout{Summing the detected probabilities $P_{\ell_B| \ell_A = 0 }$ for $\ell_A + \ell_B \equiv \text{even}$ and applying Eq. \ref{eqn:gammaextract}, we extract the Lorentz factor and verify that the measured values, $\gamma_{\text{meas}}$, correspond well to the encoded values, $\gamma$, as illustrated in Fig. \ref{fig:EXP} (c).}
As illustrated in Fig. \ref{fig:EXP} (c), we verify that the $\gamma_{\text{meas}}$ correspond well to the $\gamma$ for both methods. %However, as detailed in Table xx, the spread in OAM, $\Delta\ell$, remains constant for any fixed value of $\ell_A$, and only increases as a function of the Lorentz factor. xx more intuition/explanation will be added. 
 
From the measured $\gamma_\text{meas}$ values, we compute the corresponding rapidity $\eta$ using $\cosh(\eta) = \gamma_\text{meas}$, which is depicted in Fig. \ref{fig:EXP} (d).  Rapidity, in addition to the Lorentz factor, quantifies the hyperbolic angle by which coordinates are rotated in space-time. Additionally, the inset in Fig. \ref{fig:EXP} (d) shows that we reach simulated velocities of up to $0.99c$ in the lab. In each plot, the data points represent experimental measurements, while the solid line shows the theoretical prediction, demonstrating excellent agreement. 
% \KD{\sout{Additionally, Alice and Bob can use the measured spectra to perform a fit for the rapidity, also shown in Fig. \ref{fig:EXP}(d)}}.
%we determine the velocities at which our detectors, A and B, are travelling and
One should note that the broadening of the OAM spectrum can result in inaccuracies when extracting higher $\gamma$ values. As shown for $\gamma = 20$ in Fig. \ref{fig:EXP}(a), the spectrum extends beyond the detection range ($\ell_{A,B} = \{-20, 20\}$), suggesting that some probability amplitudes fall outside the measurable range. These unaccounted contributions can lead to discrepancies at large $\gamma$. To account for this, the data represented in \KD{Fig.}~\ref{fig:EXP}~(b-d) was computed using measured probabilities in the detection range, $\ell_{A} = 0$, $\ell_{B} = \{-40, 40\}$. Moreover, to eliminate contributions from noise at large $\ell_B$, we performed accidental subtraction and subtracted the minimum value in each OAM probability spectrum. 

%\section{Discussion and Conclusion}
%Here, we propose using a photon’s OAM spectrum to detect physical parameters from special relativity by exploiting spectral broadening in OAM due to Lorentz boosts. We show that the technique can be used to estimate the Lorentz factor, $\gamma$, from a pair of synchronized relativistic inertial reference frames. We show that in such frames, length contraction rescales spatial coordinates of detectors and distorts the orthogonality of measured entangled photons, producing a measurable broadening in detected OAM  correlations. For two photons entangled in OAM, this effect modifies their joint correlations, enabling direct inference of $\gamma$ from the altered distribution. To validate our approach, we generate entangled photon pairs via SPDC and emulate relativistic length contraction with spatial light modulators. By comparing the measured OAM spectra against theoretical predictions, we demonstrate quantitative agreement and establish structured light as a novel metrology tool for probing relativistic effects.

While this work only considered the motion of the detectors in inertial reference frames, both moving at the same velocities, it is easily extended to probe richer scenarios, such as accelerated reference frames or detectors moving at disparate speeds \cite{wu2023orbital, wu2024orbital}. Or to using photons entangled in time to photograph objects moving at relativistic speeds, extending the work of Ref. \cite{hornof2025snapshot}. These studies can potentially open the door to study OAM entanglement in gravitational fields \cite{fuentes2005alice}, especially given the rise of recent works that show the possibility of characterising the dynamics of black holes using the OAM of photons \cite{tamburini2022constraining, tamburini2011twisting} and the study of spatio-temporal effects that emerge from applying Lorentz boosts to vortex modes \cite{bliokh2012relativistic, bliokh2012spatiotemporal}. Further, our analysis and simulations are limited to two dimensions (transverse plane), however, a more concise study could explore operations on entangled photons in space-time (all 4 dimensions).

\section{Conclusion}
In conclusion, we have predicted the implications of relativistic length contraction on the joint OAM correlations measured by two spatially separated detectors and harnessed OAM dispersion as a tool to quantify the Lorentz factor. From these observations, we established that a one-to-one map exists between the Lorentz factor and the properties of the modified OAM spectrum. We tested this for different Lorentz factors, starting with the rest frame ($\gamma=1$) up to a emulated moving frame of $\gamma=20$. Our results agreed well with the theoretical predictions.

\bibliographystyle{unsrt}
\bibliography{references.bib}

\pagebreak
\widetext
\begin{center}
\textbf{\large Supplementary information: Orbital angular momentum of entangled photons as a probe for relativistic effects}
\end{center}

\setcounter{equation}{0}
\setcounter{figure}{0}
\setcounter{table}{0}
\setcounter{section}{0}
\setcounter{page}{1}
\makeatletter
\renewcommand{\theequation}{S\arabic{equation}}
\renewcommand{\thefigure}{S\arabic{figure}}
\renewcommand{\bibnumfmt}[1]{[S#1]}
\renewcommand{\citenumfont}[1]{S#1}

\section{Coordinate transformation}
The reverse Lorentz transformation maps Alice and Bob's transverse spatial coordinates as 
\begin{equation}
x = \gamma(x' + vt), \ \ y = y',
\end{equation}
when described from the perspective of a reference frame that is at rest. We also assume that Alice and Bob's clocks are synchronised because they travel with the same relativistic velocity. For this reason, we set t=0 since the detection events on Alice's and Bob's detectors occur simultaneously.  Therefore, the observer at rest can describe coordinates in Alice and Bobs moving reference frame as, $(x, y)  \rightarrow (x', y') = (\frac{x}{ \gamma}, y )$, observing them as a contraction of the spatial coordinates in the x-direction, i.e. an aberration that squashes the wavefunction in the x-direction. Therefore, since the wavefunctions are azimuthally dependent ($\phi$) which could be initially found from the relation  $\tan(\phi) = y/x$, mapped as
\begin{equation}
    \tan\phi'=\gamma(y/x)=\gamma \tan \phi,
\end{equation}
so that $\phi' = \arctan(\gamma \tan (\phi))$.

\section{Joint Probabilities}
Consider a two-photon state, entangled in OAM, in the same rest frame as observer Charlie (C)
\begin{align}
\ket{\psi} & = \frac{1}{\sqrt{N}} \sum_{\ell=0}^{N-1} \ket{\ell} \ket{-\ell}.
\end{align}

We can rewrite this state from the OAM basis into the azimuthal coordinate basis via
\begin{align}
\ket{\psi} &  = \frac{1}{2\pi\sqrt{N}}\sum_{\ell=0}^{N-1}\int_{0}^{2 \pi} \int_{0}^{2 \pi} \exp\left(i \ell \phi_1 \right)\ket{\phi_1} \ \exp\left(-i \ell \phi_2 \right)\ket{\phi_2} \ d\phi_1 \ d\phi_2,\\
& = \frac{1}{2\pi\sqrt{N}}\sum_{\ell=0}^{N-1}\int_{0}^{2 \pi} \int_{0}^{2 \pi} \exp\left(i \ell \left(\phi_1 -\phi_2\right)\right)\ket{\phi_1} \ket{\phi_2} \ d\phi_1 \ d\phi_2,\\
& = \frac{1}{2\pi\sqrt{N}}\int_{0}^{2 \pi} \int_{0}^{2 \pi}\sum_{\ell=0}^{N-1} \exp\left(i \ell \left(\phi_1 -\phi_2\right)\right)\ket{\phi_1} \ket{\phi_2} \ d\phi_1 \ d\phi_2,\\
& \approx \frac{1}{2\pi\sqrt{N}}\int_{0}^{2 \pi} \int_{0}^{2 \pi} 2\pi\delta (\phi_1 - \phi_2) \ket{\phi_1} \ket{\phi_2} \ d\phi_1 \ d\phi_2,\\
& = \frac{1}{\sqrt{N}}\int_{0}^{2 \pi}  \ket{\phi} \ket{\phi} \ d\phi.
\end{align}
This holds for large values of $N$, i.e. if there is a sufficiently large number of OAM modes that are entangled in the system. Furthermore, the identity for the Dirac Delta  function was evoked, i.e. $\sum_{\ell=0}^{N-1} \exp\left(i \ell \left(\phi_1 -\phi_2\right)\right) \propto \delta(\phi_1-\phi_2) $ . 

\subsection{Rest Frame Case}
Consider the case where Charlie sends the entangled photons to Alice (B) and Bob (B), who are at rest with respect to Charlie (a frame at rest), and Alice and Bob project the two photons onto states $\ket{\ell_A}$ and $\ket{\ell_B}$. It is assumed that the azimuthal coordinate basis has the resolution $\langle\phi_1 | \phi_2\rangle = \delta(\phi_1-\phi_2) $ and is therefore complete, which implies that inner products with the $\ket{\phi}$ modes map onto the $\phi$ dependent wavefunctions, i.e. $\langle \Psi | \phi \rangle = \langle \phi | \Psi \rangle^*   = \Psi^*(\phi)$. In this scenario, Alice and Bob's measurements can be described as 
\begin{align}
    \ket{\ell}_{A,B} = \frac{1}{\sqrt{2\pi}}\int_0^{2\pi}\exp\left(i \ell_{A,B}\phi_3\right)\ket{\phi_3}\ d\phi_3,
\end{align}
where $\langle \phi_3 |  \ell_{A, B} \rangle = \exp\left(i \ell_{A,B}\phi_3\right)$.

Then, the joint probabilities are given by
\begin{align}
P_{\ell_{A}\ell_B} &= \left|\langle{\ell_A}|\langle{\ell_B |\psi} \rangle \right|^2\\
&= \frac{1}{(2\pi)^2N}\left| \int_{0}^{2 \pi}\int_{0}^{2 \pi}\int_{0}^{2 \pi} \exp\left( - i \ell_{A} \phi_3 \right) \bra{\phi_3} \ \exp\left( - i \ell_{B} \phi_4 \right) \bra{\phi_4} \ket{\phi_1} \ket{\phi_1} \ d\phi_1 \ d\phi_3 \ d\phi_4 \right|^2 \\
&= \frac{1}{(2\pi)^2N}\left|  \int_{0}^{2 \pi} \int_{0}^{2 \pi}\int_{0}^{2 \pi} \exp\left( - i \ell_{A} \phi_3 \right)  \exp\left( - i \ell_{B} \phi_4 \right) \delta\left(\phi_1 - \phi_3\right)\delta\left(\phi_1 - \phi_4\right)  \ d\phi_1 \ d\phi_3 \ d\phi_4 \right|^2 \\
&= \frac{1}{(2\pi)^2N}\left|  \int_{0}^{2 \pi} \exp\left( - i \left(\ell_{A} + \ell_{B}\right) \phi_1 \right)\ d\phi_1 \right|^2 \\
&= \frac{1}{(2\pi)^2N}\left| 2\pi\delta_{\ell_A , - \ell_B}  \right|^2 \\
&= \frac{\delta_{\ell_A , - \ell_B}}{N}
\end{align}

\subsection{Relativistic Case}
Now consider the case where Charlie instead sends the two photons to Alice and Bob, who are moving with relativistic velocities in a transverse direction with respect to Charlie. Alice and Bob want to perform projective OAM measurements on their photons, obtaining the joint overlap

\begin{align}
    \langle\ell_A|\langle\ell_B|\psi\rangle &= \frac{1}{\sqrt{N}}\bra{\ell_A}\bra{\ell_B}\int_{0}^{2 \pi}  \ket{\phi_1} \ket{\phi_1} \ d\phi_1\\
    &= \frac{1}{\sqrt{N}}\int_{0}^{2 \pi} \langle \ell_A | \phi_1 \rangle \langle \ell_B | \phi_1 \rangle \ d\phi_1,
    \end{align}
In this case, Alice and Bob's measurements have to account for relativistic effects. But, since the joint probability depends on the projection using the $\phi$ modes that are defined using the coordinate system in the reference that's at rest (Charlie), we can also describe Alice and Bob states using the same basis since the two photon states essentially project onto the corresponding $\phi$ dependent wavefunctions so that 
\begin{align}
    \langle \ell_{A, B} | \phi_1 \rangle &= \frac{1}{\sqrt{2\pi}}\int_0^{2\pi}\exp\left(-i \ell_{A,B} \phi'\left(\phi_3\right)\right) \langle \phi_3|\phi_1\rangle  d\phi_3\\
    &= \frac{1}{\sqrt{2\pi}}\int_0^{2\pi}\exp\left(-i \ell_{A,B} \phi'\left(\phi_3\right)\right)\delta \left(\phi_3 - \phi_1\right)\ d\phi_3\\
    &= \frac{1}{\sqrt{2\pi}}\exp\left(-i \ell_{A,B} \phi'\left(\phi_1\right)\right),
\end{align}
and there we can now simplify the integral as 
\begin{align}
      \langle\ell_A|\langle\ell_B|\psi\rangle  &=  \frac{1}{\sqrt{N}}\int_{0}^{2 \pi} \langle \ell_A | \phi_1 \rangle \langle \ell_B | \phi_1 \rangle  d\phi_1, \\
 &= \frac{1}{2\pi\sqrt{N}}\int_{0}^{2 \pi} \exp\left(-i \ell_A \phi'\left(\phi_1\right)\right)\exp\left(-i \ell_B \phi'\left(\phi_1\right)\right) \ d\phi_1\\
    &= \frac{1}{2\pi\sqrt{N}}\int_{0}^{2 \pi} \exp\left(-i \left(\ell_A +\ell_B\right)\phi'\left(\phi_1\right)\right) \ d\phi_1.
\end{align}

Hence, the joint probability, $P_{\ell_A \ell_B}$, for Alice and Bob projecting the $\ket{\psi}$ state onto $\ket{\ell_A}$ and $\ket{\ell_B}$ is given by
\begin{align}
    P_{\ell_A\ell_B} & =  \left|\langle\ell_A|\langle\ell_B|\psi\rangle\right|^2\\
    &= \left| \frac{1}{2\pi\sqrt{N}}\int_{0}^{2 \pi} \exp\left(-i \left(\ell_A +\ell_B\right)\phi'\left(\phi_1\right)\right) \ d\phi_1 \right|^2.
\end{align}

These joint probabilities will be proportional to the coincidence counts that they obtain experimentally.  Charlie describes his reference frame by spatial coordinates ($x$, $y$), and so observes his azimuthal coordinate $\phi$ as $\tan\phi = \frac{x}{y}$. For another reference frame experiencing a Lorentz boost along $x$, Charlie observes that reference frame's coordinates as length contracted, namely  ($x'$, $y'$) = ($\frac{x}{\gamma}$, $y$), and observes that reference frame's azimuthal coordinate $\phi'$ as $\tan\phi' = \frac{y}{\frac{x}{\gamma}} = \gamma \tan\phi$.
Hence
\begin{align}
    \phi&=\arctan(\frac{1}{\gamma}\tan\phi'),
\end{align}
To calculate the Jacobian of this coordinate transformation, substitute $u(\phi') = \frac{1}{\gamma}\tan\phi'$. Then
\begin{align}
    \frac{d\phi}{d\phi'} &= \frac{d\arctan u}{du}\cdot\frac{du}{d\phi'}\\
    &= \frac{1}{1+u^2}\cdot\frac{1}{\gamma \cos^2 \phi'}\\
    &= \frac{1}{1+\frac{1}{\gamma^2}\tan^2\phi'}\cdot\frac{1}{\gamma \cos^2 \phi'}\\
    &= \frac{\gamma^2}{\gamma^2 + \frac{\sin^2\phi'}{\cos^2\phi'}}\cdot\frac{1}{\gamma \cos^2 \phi'}\\
    &= \frac{\gamma}{\gamma^2\cos^2 \phi' + \sin^2\phi'}\\
    &= \frac{\gamma}{\gamma^2\cos^2 \phi' + 1 - \cos^2\phi'}\\
    &= \frac{\gamma}{(\gamma^2-1)\cos^2 \phi' + 1'}.
\end{align}
Therefore
\begin{equation}
    d\phi = \frac{\gamma}{(\gamma^2-1)\cos^2 \phi' + 1'} d\phi'.
\end{equation}
The joint probabilities can now be mapped onto 

\begin{equation}
   P_{\ell_A \ell_{B}} =  \left | \frac{1}{2\pi \sqrt{N}} \int_{0}^{2 \pi}  \frac{\gamma e^{-i(\ell_A + \ell_B) \phi'(\phi)}}{(\gamma^2-1)\cos^2(\phi') +1}   d \phi' \right |^2.
\end{equation}\\
Subsequently, we apply the residue theorem to the integral after performing the mapping $z = \exp(-i \phi')$, $\cos(2\phi') = (z^2 + z^{-2})/2$, and $d \phi' = -1/(iz) \ dz$ so that 
\begin{equation}
   P_{\ell_A \ell_{B}} =  \left | \frac{1}{\sqrt{N}} \int_{|z| = 1}    \frac{2 i \gamma  z^{\ell_A+\ell_B+1} d z }{\pi \left(\gamma +\gamma  z^2-z^2+1\right) \left(\gamma +\gamma  z^2+z^2-1\right)}   \right |^2.
\end{equation}
Noting that the new function mapping has poles at $z_{\pm} = \pm\frac{\sqrt{1-\gamma }}{\sqrt{\gamma +1}}$ that are within the boundary $|z| = 1$, we can evaluate the integral using the residue theorem, i.e. $\int_{|z|=1} f(z) dz =  2 \pi i\sum_k \text{Res}(f, z_k)$. Evaluation using the above-mentioned poles ($z_{\pm}$), results in the closed form elementary solution
\begin{equation}
   P_{\ell_A \ell_B} =  \frac{1}{N}  \left( \frac{\gamma -1 }{\gamma+1} \right)^{|\ell_{A}+\ell_{B}|} \cos^2((\ell_A+\ell_B)\frac{\pi}{2}), 
\end{equation}
\\
showing that the spectra is defined when  $\ell_{A}+\ell_{B}$ is even. 

Further, the function  $\left( \frac{\gamma -1 }{\gamma+1} \right)^{|\ell|}$  (with $\ell = \ell_{A}+\ell_B$) is a simple geometric sequence resulting in the convergent series $\sum^{\infty}_{l=-\infty} \left( \frac{\gamma -1 }{\gamma+1} \right)^{|l|} =  \gamma$.

\end{document}